\documentclass{aa}  
\usepackage{booktabs}
\usepackage{comment}
\usepackage{xcolor}
\usepackage{xspace}
\usepackage{graphicx}
\usepackage{txfonts}
\usepackage[nolist]{acronym}
\usepackage{hyperref}

\newcommand{\mghk}{\ion{Mg}{II} h\&k\xspace}
\newcommand{\mgh}{\ion{Mg}{II} h\xspace}
\newcommand{\mgk}{\ion{Mg}{II} k\xspace}

\newcommand{\caefft}{\ion{Ca}{ii}~$\lambda854.2$ nm\xspace}

\newcommand*{\ie}{i.e.\@\xspace}
\newcommand*{\eg}{e.g.\@\xspace}

\newcommand{\konev}{$\mathrm{k_{1v}}$\xspace}
\newcommand{\koner}{$\mathrm{k_{1r}}$\xspace}
\newcommand{\kone}{$\mathrm{k_{1}}$\xspace}
\newcommand{\ktwor}{$\mathrm{k_{2r}}$\xspace}
\newcommand{\ktwov}{$\mathrm{k_{2v}}$\xspace}
\newcommand{\ktwo}{$\mathrm{k_{2}}$\xspace}
\newcommand{\kthree}{$\mathrm{k_{3}}$\xspace}

\newcommand{\hone}{$\mathrm{h_{1}}$\xspace}

\newcommand{\htwo}{$\mathrm{h_{2}}$\xspace}
\newcommand{\hthree}{$\mathrm{h_{3}}$\xspace}

\newcommand{\cahk}{ Ca~{\sc ii}~H \& K\xspace}

\newcommand{\iris}{ IRIS\xspace}

\newcommand{\km}{\ensuremath{\, \mathrm{km}}\xspace}

\newcommand{\G}{\ensuremath{\, \mathrm{G}}\xspace}
\newcommand{\tauUnity}{\ensuremath{\tau_{500}=1}\xspace}
\newcommand{\kms}{\ensuremath{\, \mathrm{km\,s^{-1}} }\xspace}
\newcommand{\Mm}{\ensuremath{\, \mathrm{Mm}}\xspace}

\newcommand{\be}{\begin{equation}}
\newcommand{\ee}{\end{equation}}

\def\h2{\ensuremath{\mathrm{H}_2}}

\usepackage[normalem]{ulem}

\newcommand{\snapfourNineNine}{\texttt{muram\_en\_499000\_379s}\xspace}

\defcitealias{2024A&A...692A...6O}{Pub~I}

\begin{document}

\begin{acronym}

\acro{lte}[LTE]{local thermodynamic equilibrium}
\acro{rte}[RTE]{radiative transfer equation}
\acro{te}[TE]{thermodynamic equilibrium}
\acro{ne}[NE]{nonequilibrium}

\acro{nlte}[NLTE]{non-LTE}
\acro{se}[SE]{statistical equilibrium}
\acro{los}[LOS]{line-of-sight}
\acro{eos}[EoS]{equation of state}

\acro{crd}[CRD]{complete frequency redistribution}
\acro{prd}[PRD]{partial frequency redistribution}
\acro{rh15d}[RH1.5D]{Rybicki \& Hummer 1.5D RT code}
\acro{rt}[RT]{radiative transfer}
\acro{mhd}[MHD]{magnetohydrodynamics}
\acro{rmhd}[rMHD]{radiaton-MHD}
\acro{iris}[IRIS]{Interface Region Imaging Spectrometer }
\acro{muram}[MURaM]{MURaM}
\acro{muramche}[MURaM-ChE]{chromospheric extension of MURaM}
\acro{clv}[CLV]{center-to-limb variation}
\acro{euv}[EUV]{extreme ultra violet}
\acro{nir}[NIR]{near-infrared}
\acro{nuv}[NUV]{near-ultraviolet}
\acro{uv}[UV]{ultraviolet}
\acro{ali}[ALI]{approximate Lambda iteration}
\acro{ff}[ff]{free-free}
\acro{qs}[QS]{quiet Sun}
\acro{soup}[SOUP]{Solar Optical Universal Polimeter}
\acro{sst}[SST]{Swedish $1 \, \mathrm{m}$ Solar Telescope}
\acro{chromis}[CHROMIS]{CHROMospheric Imaging Spectrometer}
\acro{gong}[GONG]{Global Oscillation Network Group}
\acro{dot}[DOT]{Dutch open telescope}
\acro{ar}[AR]{active regions}
\acro{fb}[fb]{free-bound}
\acro{bf}[bf]{bound-free}
\acro{en}[EN]{enhanced network}
\acro{fov}[FOV]{field-of-view}
\acro{roi}[ROI]{region of interest}
\acro{bb}[bb]{bound-bound}
\acro{fwhm}[FWHM]{full width at half maximum}
\acro{ssd}[SSD]{small-scale dynamo}
\acro{fts}[FTS]{Fourier Transform Spectrograph}
\acro{gl}[GL]{Gau\ss-Legendre}
\end{acronym}

\title{\mghk spectral line properties computed using 3D radiative transfer in an enhanced network region simulated with the MURaM-ChE code}

\author{ P. Ondratschek\inst{1}, D. Przybylski\inst{1}, H.N. Smitha\inst{1}, J. Leenaarts\inst{2}, R. Cameron\inst{1}, S.K. Solanki\inst{1}}
\authorrunning{Ondratschek et al.}
\titlerunning{\mghk spectra, 3D RT results}

\institute{Max-Planck-Institut f\"ur Sonnensystemforschung, Justus-von-Liebig-Weg 3, 37077 G\"ottingen, Germany
            \mail{ondratschek@mps.mpg.de}
            \and Institute for Solar Physics, Department of Astronomy, Stockholm University, AlbaNova University Centre, 106 91 Stockholm,
Sweden
          }

\abstract
{
Synthetic spectral lines from radiation-magnetohydrodynamic models are essential to interpret the observations of the solar chromosphere, but only a few computations in the literature take 3D effects in the synthesis into account.  The \mghk lines form in the middle to upper chromosphere and are well-suited to study the structure of the chromosphere. However, the details of their formation in the solar chromosphere are not fully understood. 
}
{
We aim to study the effects of 3D \ac{rt} on the \mghk line properties and to verify known correlations between the underlying atmosphere and spectral line features in a new model of the chromosphere.}
{
We forward model the \mghk lines in 3D \ac{rt} with partial frequency redistribution (PRD) in a self-consistent 3D radiative magnetohydrodynamics (rMHD) simulation with non-local-thermodynamic-equilibrium (NLTE) energy transport and non-equilibrium (NE) hydrogen ionization of an \ac{en} region simulated with the \ac{muramche}.  
 We compare the spectra computed with 3D RT to those computed with 1.5D RT and to observations from the Interface Region Imaging Spectrograph (IRIS). Furthermore, we test correlations between the spectral line properties and underlying atmospheric properties such as temperature and velocity structure. 
}
{
The spatially averaged \mghk spectral lines computed with 3D RT match approximately a typical IRIS observation, which includes quiet sun and network elements. The peak separation, is however, still slightly lower in the simulation. In contrast, the 1.5D RT spectra tend to overestimate the peak intensities and the central minimum significantly. In the \ac{muramche} model, the qualitative difference between 1.5D and 3D RT results is even more pronounced than in the public Bifrost snapshot, as given in the literature. We found that this large discrepancy might partly be attributed to the horizontal velocities that are naturally included in the full 3D \ac{rt} synthesis but not in typical 1.5D \ac{rt} computations. We confirm that correlations between spectral line properties and the underlying atmosphere from the MURaM-ChE simulation are similar to those obtained from Bifrost, but show more scatter due to the more dynamic atmosphere. In addition to already known correlations, we found that the \ktwov (blue) peak of the \mgk line forms preferably in upflows whereas the \ktwor (red) peak forms preferably in downflows.}
 {The \mghk lines computed with 3D RT match the observations better in the core intensities and their distribution on the Sun compared to 1.5D computations. This underlines the importance of 3D RT in the forward modeling of \mghk.}

   \keywords{Sun - chromosphere, Sun - atmosphere, magnetohydrodynamics (MHD), radiation transfer (RT)}

   \maketitle
   \nolinenumbers

\section{Introduction} 
The solar chromosphere is an interface between the underlying photosphere and the corona above. The causes of many phenomena such as the heating of the chromosphere \citep[\eg,][]{1977ARA&A..15..363W}, or the strong dynamics it displays are not understood. Additionally, the evolution of fine structures such as spicules, their interaction with the magnetic field, and their role in supplying mass and energy to the corona are under debate \citep{2019ARA&A..57..189C}. Inference of physical quantities such as temperature, velocity, density, and magnetic field is limited through a few strong spectral lines forming in the chromosphere. Because of the decreasing density with height, spectral line formation in the chromosphere is subject to \ac{nlte} conditions. In addition, some strong resonance lines such as \cahk and \mghk show \ac{prd} effects through scattering \citep[\eg][]{1974ApJ...192..769M}. Furthermore, already in the 1980s it became clear from Lyman $\alpha$ observations that the chromosphere is inhomogeneous and plane-parallel modeling is only a zeroth-order approximation \citep[\eg,][]{1980ApJ...237L..47B}. Only recently has it become possible to model chromospheric lines including both, a 3D model atmosphere and 3D \ac{rt}. Examples are the \caefft line \citep{2009ApJ...694L.128L}, H$\alpha$ \citep{2012ApJ...749..136L}, \mghk \citep{2013ApJ...772...89L} and \citep[with a \ac{prd} treatment,][]{2017A&A...597A..46S}, and \cahk \citep[][]{2018A&A...611A..62B}. An alternative to the computationally expensive 3D \ac{rt} computations is 1.5D \ac{rt}. In this approximation, the spectral lines are computed from a 3D atmosphere model, but each vertical column is treated as an independent plane-parallel atmosphere.

Two main strategies are used to interpret observations.  Inversions aim to reconstruct an atmosphere model to fit an observed spectral line profile. However, the model atmospheres are kept relatively simple, and the solution is not always unique. In other words, different model atmospheres may lead to very similar spectral line shapes \citep[for reviews of inversions see eg,][]{2016LRSP...13....4D,2017SSRv..210..109D}. In addition, fudge parameters such as microturbulence are introduced to fit the observed line widths, while the actual velocity structure remains unresolved. Inversion codes solve the \ac{rt} problem, but they are generally restricted to 1.5D \ac{rt} as their calculations can easily become expensive, although approaches to take 3D \ac{rt} effects in inversions into account exist \citep{2022A&A...659A.137S}.

The other main strategy is based on forward modeling. In the first step, a model of the solar atmosphere is simulated. In the second step, detailed \ac{rt} computations are conducted post-processing. In contrast to inversions, forward modeling does not aim to exactly reproduce a single observation. The advantage is, however, that through the knowledge of all quantities in the atmosphere, line formation can be studied in detail. By comparing the atmosphere with the resulting spectra, possible correlations can be identified, which can be used to interpret observations without performing inversions.

In this paper, we concern ourselves with the \mghk\ lines. A typical line profile of \mgk observed in the \ac{qs} shows a double-reversed line core. The inner wings, which are the minimum intensities before the first reversal are called \konev and \koner, where the ``v'' indicates the blue or violet and ``r'' the red part of the spectrum with respect to the rest wavelength. The peaks are similarly labeled \ktwov and \ktwor. The central reversal is called \kthree. The \mgh line shows similar properties as \mgk and has therefore analogous labeling. These spectral features form at different atmospheric heights \citep[see \eg,][Fig.~1]{1981ApJS...45..635V} and are thus sensitive to local conditions. The \kone features form in the low chromosphere close to the temperature minimum in a plane-parallel atmosphere. The \ktwo peaks form in the mid chromosphere and the \kthree minimum just below the transition region.

Recent studies of the \mghk line formation were presented by \citet{2013ApJ...779..155A} for different regions on the sun using 1D model atmospheres.  The authors investigated the \ac{clv} and the effect of up and downflows on the resulting \mgk line profiles. \citet{2013ApJ...772...89L,2013ApJ...772...90L} and \citet{2013ApJ...778..143P} used a snapshot of a 3D \ac{en} simulation \citep[][herafter public Bifrost snapshot]{2016A&A...585A...4C} computed with the Bifrost code \citep{2011A&A...531A.154G}. There it was found, for example, that the Doppler shifts of the \kthree and \hthree features are good indicators of the vertical velocity in the upper chromosphere, while differences between the \kthree and \hthree Doppler shifts correlate with velocity gradients in the upper chromosphere. The \ktwo and \htwo peak intensities can be used to estimate the temperature in the middle chromosphere, while the peak intensity ratios of the \ktwo or \htwo peaks correlate with up- or downflows in the upper chromosphere. A summary of these diagnostics can be found in \citet[][Tab.~4]{2013ApJ...778..143P}. These correlations have been proven useful, among other studies, to interpret variability of \mgh in full disk mosaics by \citet{2015ApJ...811..127S}, to constrain chromospheric dynamics by \citet{2018ApJ...857...48G} and \citet{2019ApJ...881..109H}, and to study the velocity structure of prominences by \citep[\eg,][]{2021A&A...653A...5P}. A phenomenon of the \mghk line formation that has so far not been entirely understood is the difference in the emergent intensity between the \ac{qs} and coronal holes \citep{2018ApJ...864...21K}.

The computations of \citet{2013ApJ...772...90L} and \citet{2013ApJ...778..143P} resulted in \mghk profiles with too-narrow line widths and too-weak peak intensities when compared to observations. The above authors and \citet{2016A&A...585A...4C} argued that a higher numerical resolution in the simulation might help to resolve more dynamics and reduce such discrepancies. Indeed, the simulations of \citet{2023ApJ...943L..14M}  showed a better match with observed profiles by using a high horizontal resolution of $5 \km$. The flux emergence simulations of \citet{2023ApJ...944..131H} also show a good match with observations, even though the spatial resolution of $100 \km$ is relatively low. This suggests higher mass loading of the chromosphere is required to reproduce the observed line width. Recently, \cite{2024A&A...692A...6O}, hereafter \citetalias{2024A&A...692A...6O}, presented a relatively close match of the \mghk lines from an \ac{en} simulation with observations of the \ac{qs}. The simulation they used was computed with the recently developed \ac{muramche}\footnote{Max Planck Institute for Solar System Research/University of
Chicago Radiation Magneto-hydrodynamics with the chromospheric extension.} code \citep{2022A&A...664A..91P}. While the resolution in the \ac{muramche} model was relatively moderate, that is $20\km$ vertically and $23.4\km$ horizontally, in \citetalias{2024A&A...692A...6O} it was found that the \ac{muramche} atmosphere is more dynamic than the public Bifrost snapshot, explaining the improved match.

The authors also found that the \mghk peak intensities are larger and the line width slightly smaller in the spatially averaged simulated profiles than in the observations. In \citetalias{2024A&A...692A...6O} the \mghk spectra were computed in the 1.5D \ac{rt} approximation where each column in the atmosphere is treated as an independent plane-parallel atmosphere. 

In this work, we aim to extend these studies by computing the spectra of the same simulation snapshot but in 3D \ac{rt} by using the Multi3D code \citep{2009ApJ...694L.128L}. Our goal is to study the impact of 3D \ac{rt} for the \mghk lines in the \ac{muramche} model, which was shown to be important by \citet{2017A&A...597A..46S}, \citet{2019A&A...631A..33B}, and \citet{2020ApJ...901...32J}. In addition, we aim to test the robustness of the correlations found by \citet{2013ApJ...772...90L} and \citet{2013ApJ...778..143P} in the \ac{muramche} simulation used here.

This paper is structured as follows. In Section \ref{sec:methods}, we briefly describe the MURaM-ChE model, the RT computations, and the observation we use for comparison. In Section \ref{sec:results}, we describe our results, and in Section \ref{sec:summary_and_conclusion}, we provide a summary and discussion. In section \ref{sec:conclusion}, we present our conclusions.

\section{Methods}
\label{sec:methods}
\begin{figure*}
\centering
\sidecaption
\includegraphics[width=12cm]{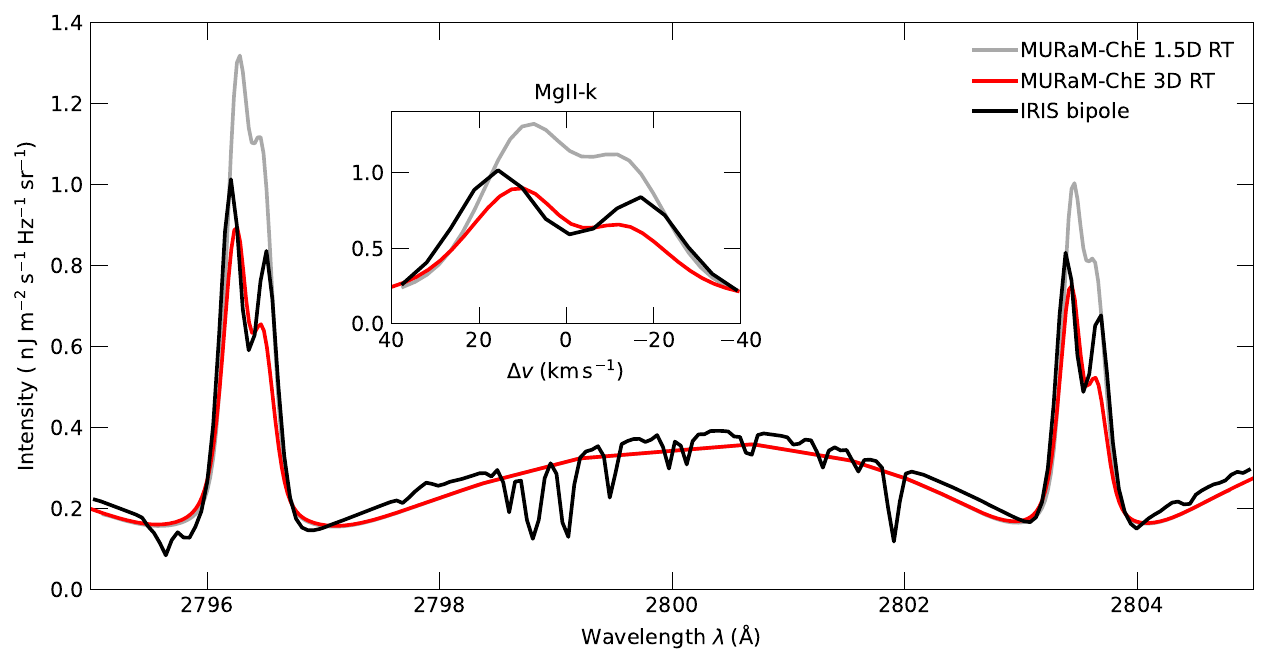}
  \caption{Spatially averaged spectra of the \mghk lines. Shown are the spectra from the \ac{muramche} simulation once computed in full 3D \ac{rt} (red) and once using the plane-parallel 1.5D \ac{rt} approximation (grey). For comparison, we show an observation from a qualitatively similar bipolar feature, which covers a similar area on the Sun (black). }
     \label{fig:average_spectrum}
\end{figure*}

In the following, we describe the \ac{muramche} \ac{en} simulation and the Multi3D code, which we used to compute the \mghk spectra in 1.5D \ac{rt} and 3D \ac{rt}. We then describe how the spectra were degraded to the instrumental conditions of the \ac{iris}. After this, we describe the observation to which we compare our results.

\subsection{The MURaM-ChE model}
We use the same model atmosphere as described in \citetalias{2024A&A...692A...6O}, which represents an \ac{en} region with a large-scale bipolar structure. This model was designed to be similar to the public Bifrost snapshot \citep{2016A&A...585A...4C}, including the same large-scale field geometry. The model is simulated with the chromospheric extension of the MURaM \ac{rmhd} code \citep{2022A&A...664A..91P}, which includes \ac{nlte} line losses and a \ac{ne} treatment of hydrogen ionization. The model atmosphere extends over a solar volume of $24 \Mm \times 24 \Mm \times 24 \Mm$ with a horizontal resolution of $23.4 \km$ and a vertical resolution of $20 \km$. The \ac{en} simulation is based on the \ac{ssd} simulation by \citet{2025A&A...703A.148P}. The convection zone extends roughly from $z=-7 \Mm$ to $z=0 \Mm$, where $z=0 \Mm$ lies at the average \tauUnity height. The atmosphere extends up to $z=17 \Mm$. We refer to \citetalias{2024A&A...692A...6O} for more details.

\subsection{The Multi3D code}

To compute the spectra of \mghk in 3D \ac{rt} we utilize the Multi3D code \citep{2009ASPC..415...87L} with the extension to compute \ac{prd} spectra \citep{2017A&A...597A..46S}. In Multi3D, the \ac{rte} and the equations of \ac{se} are solved simultaneously. The solution is computed iteratively based on the multilevel accelerated $\Lambda$-iteration (MALI) scheme with preconditioned radiative rates according to \citep{1991A&A...245..171R, 1992A&A...262..209R} until convergence of relative population changes with a tolerance of $10^{-3}$ is reached. The \ac{rte} is integrated by using the short characteristics scheme. This allows for domain decomposition and a parallel treatment of the problem. The anglular quadrature follows the 24-angle “A4'' set of \citet{carlson_1963}.
We compute the \mghk lines by using the 4+1 level model atom similar to \citet{2017A&A...597A..46S} and \citet{2019A&A...631A..33B}. The model atom is described in \citet{2013ApJ...772...89L}. This is a compromise between accuracy and computational time. As demonstrated by \citet{2013ApJ...772...89L} the 4+1 level shows only small deviations from the more accurate 10+1 level atom and is therefore a good choice for expensive 3D \ac{rt} computations.
We used only every other column from the 3D MHD cubes to speed up the computation. This effectively reduces the horizontal resolution to $\approx 48 \km$.  Multi3D has the option to compute spectra also in the 1.5D \ac{rt} plane-parallel mode. We used this mode to compare between the 1.5D and 3D \ac{rt} results. This has the advantage that differences between separate codes can be avoided. Unlike the computation in \citetalias{2024A&A...692A...6O} we did not include blend lines.

\subsection{Degrading spectra to instrumental conditions}
To compare the synthetic intensities with the observation from IRIS, we degraded the spectra to the instrumental conditions as described in \citet{2014SoPh..289.2733D}. We performed the same procedure as described in \citetalias{2024A&A...692A...6O} based on the description in \citet{2013ApJ...778..143P}. We convolved the spectra spatially with a Gaussian kernel of $0.4"$ \ac{fwhm}. In addition, the spectral profiles are convolved with a Gaussian kernel of $6 \, \mathrm{pm}$ \ac{fwhm}. The spectra are then rebinned to a spatial grid of $0.16" \times 0.33"$ pixel size, which results in a final resolution of $199 \times 100$ pixels.

\subsection{Observation}

We compare our results to an \ac{iris} observation of a similarly sized \ac{fov} as the simulation of $24 \Mm \times 24 \Mm$ or $33" \times 33"$. This is a subregion of a very large dense raster observation of $141" \times 175"$ that is located close to disc center at the heliocentric coordinates $(x,y) = (49",2")$. The observation was taken on 2014-06-02. The unsigned mean of the \ac{los} magnetic field, obtained from the Helioseismic and Magnetic Imager \citep[HMI,][]{2012SoPh..275..207S,2012SoPh..275..229S}, in the subregion is $10.76 \G$, which compares to the observation that we used in \citetalias{2024A&A...692A...6O}. We chose this observation as it shows more fine structure in the line center than the one used in \citetalias{2024A&A...692A...6O}. We note that this serves only as a rough comparison, as the setup of the numerical simulation is not meant to reproduce this observation. A discussion on the variation of the \mghk spectra depending on the observed region can for example be found in \citetalias[][Appendix~A]{2024A&A...692A...6O}. In Sect.~\ref{sec:intensity_images_observation_and_synthesis}, we present a qualitative comparison between the observation and the synthetic images computed from the \ac{muramche} model.

\begin{figure*}
\sidecaption
\centering
\includegraphics[width=12cm]{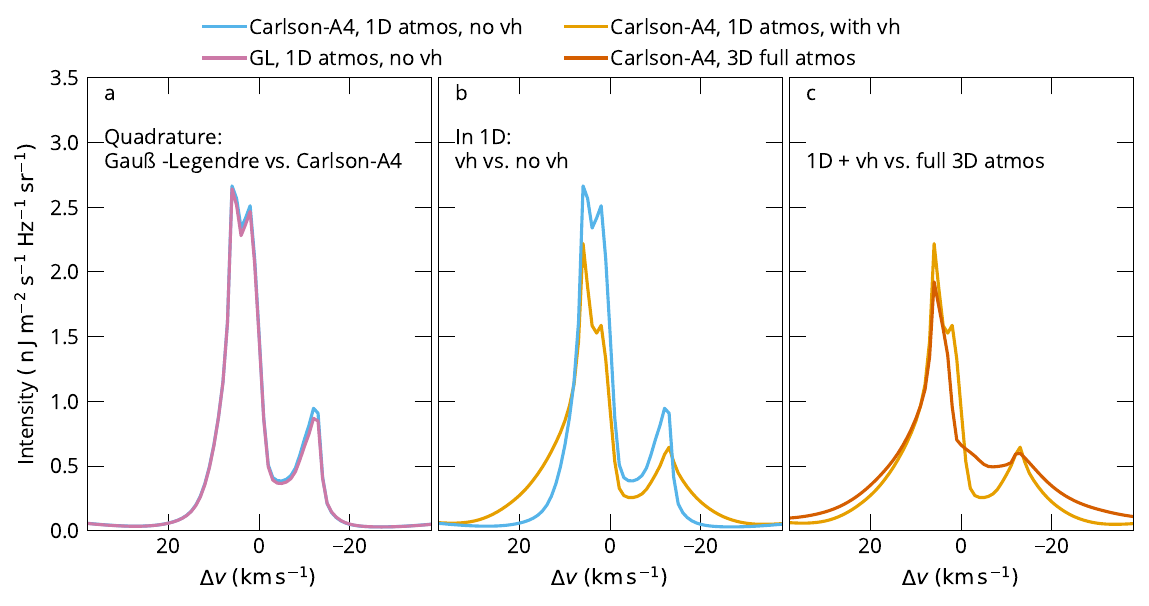}
\caption{The effect on the spectral line profile of varying the angular quadrature, and of including horizontal velocities ''vh`` in the \ac{rt} computation. The figure shows profiles calculated from a single pixel of the simulation. In panel (a), we compare spectra computed assuming a plane-parallel atmosphere, neglecting horizontal velocities, with two different angular quadratures: the Gau\ss-Legendre (GL) quadrature and the \citet{carlson_1963}-A4 quadrature. In panel (b), we compare the computations from the plane-parallel atmosphere without horizontal velocities with a computation that takes the horizontal velocities into account. In panel (c), we compare the computed spectrum from the 1D plane-parallel atmosphere with horizontal velocities with the spectrum computed in full 3D at the same location. }
 \label{fig:fig-quadrature-and-vh}
\end{figure*}

\section{Results}

\label{sec:results}
Our analysis consists of two parts. First, we compare the spatially averaged spectrum from the simulation with a similarly averaged 
observed spectrum. In addition, we compare intensity images from the 1.5D and 3D \ac{rt} synthesis as well as statistical distributions of spectral line properties with the observed data. For this analysis, we use synthetic spectra that are degraded to \ac{iris} instrumental conditions. In the second part of the results, we test based on the 3D \ac{rt} results whether similar correlations between the spectra and the underlying atmosphere exist, as found by \citet{2013ApJ...772...90L} and \citet{2013ApJ...778..143P}. We focus on the \mgk line as the formation of the \mgh line is similar. For this analysis we use the spectra at the original computed spatial and spectral resolution. This preserves the information about the wavelength-dependent optical depth that we use to estimate the formation height in the atmosphere. The peak features are identified by the method described in \citetalias[][]{2024A&A...692A...6O} which is based on the method by \citet{2013ApJ...778..143P}.

\subsection{Comparison with observation and 1.5D \ac{rt} computation}
Here we present comparisons between the forward-modeled spectra and the observation. We start by comparing the spatially averaged spectrum. We then compare intensity images at the rest wavelength and the spectral line features. Finally, we compare statistical distributions of the \ktwo peak intensity, peak separation, and peak intensity ratio.

\subsubsection{Average spectrum}
\label{sec:results_average_spectra}
In Fig.~\ref{fig:average_spectrum}, we present the spatially averaged spectrum of a region observed by \ac{iris}, the spectrum from the simulation computed with the 1.5D \ac{rt} approximation, and the spectrum calculated from the same snapshot but with 3D \ac{rt}. It can be seen that the intensity of the synthetic spectra matches the observation in the \kone minima and the pseudo-continuum in between the \mgk and \mgh line with the observations. The mismatch with the spectral lines between the two \mghk lines is due to not including such blending lines in the computations with Multi3D. 

In the \ktwo peaks and \kthree minimum, the 1.5D and 3D \ac{rt} computations differ. The intensity of the \ktwov peak from the 3D average spectrum is $32\%$ lower than the intensity of the \ktwov peak from the 1.5D average spectrum. Similarly, the \ktwor peak intensity in 3D \ac{rt} is $41 \%$ lower than in the 1.5D \ac{rt} computation. The central minimum in 3D \ac{rt} is $42\%$ lower than in the 1.5D \ac{rt} computation. These values are almost a factor of four higher than the values we estimated in \citetalias{2024A&A...692A...6O} from the difference in the Bifrost public snapshot between 1.5D and 3D \ac{rt} computations \citep[see][Fig.~10]{2017A&A...597A..46S}. This indicates that in the \ac{muramche} simulation, 3D \ac{rt} effects are more important as they decrease the intensity contrast. In the \kone and \hone features, and in the pseudo-continuum between the \mghk lines, the 3D \ac{rt} and 1.5D \ac{rt} computations agree with each other. While locally there can exist differences at these intensities, there seems to be no net effect when the spectrum is averaged over the whole cube. This is expected as these intensities form in the lower atmosphere, where 3D \ac{rt} is less important. We further discuss the discrepancy between the 1.5D and 3D \ac{rt} results in Sect.~\ref{sec:the_role_of_horizontal_velocities}.

In comparison with the observed spectrum, the 3D \ac{rt} computation results in a closer match of the line core, however the \ktwo and \htwo peak intensities are lower in the 3D \ac{rt} computation. The \kthree and \hthree intensity approximately matches the observed value.
The \kthree minima of the synthetic spatially averaged spectra are shifted by $-5.3 \kms$ (3D RT) and $-5.6 \kms$ (1.5D RT) with respect to the rest wavelength of \mgk, whereas in the observation the \kthree feature is shifted by $-2.3 \kms$. These values were determined through a parabolic fit around the \kthree feature of the spatially averaged profile, that is they are not the average of \kthree wavelength positions from single columns. We find that at the typical heights where the \kthree features form in the atmosphere, the average velocity is a downflow, explaining the net red shift of the spatially averaged spectrum. We discuss these results in more detail in Sect. \ref{sec:correlation_doppler_shifts_and_atmospheric_velocities}.
The peak intensity ratios computed via 
\begin{align}
    R_\mathrm{k} = (I_{\mathrm{k2v}} - I_{\mathrm{k2r}}) / (I_{\mathrm{k2v}} + I_{\mathrm{k2r}})
    \label{eq:paper-ii-peak-intensity-ratio}
\end{align} \citep[see \eg,][Eq.~2]{2013ApJ...772...90L} are $0.08$ (1.5D \ac{rt}), $0.16$ (3D \ac{rt}), and $0.09$ (\ac{iris}). This means for all three spectra the blue peak ($I_{\mathrm{k2v}}$) is stronger than the red peak ($I_{\mathrm{k2r}}$). The higher peak intensity ratio in the 3D \ac{rt} computation than in the 1.5D \ac{rt} computation is a result of the lower \ktwo intensities in 3D \ac{rt} while the difference between two \ktwo peaks, that is $I_{\mathrm{k2v}} - I_{\mathrm{k2r}}$ is similar for the two computations, that is $0.24 \,\mathrm{nJ\,m^{-2}\,s^{-1}\,Hz^{-1}\,sr^{-1}}$ (3D \ac{rt}) and $0.2 \,\mathrm{nJ\,m^{-2}\,s^{-1}\,Hz^{-1}\,sr^{-1}}$ (1.5D \ac{rt}). The peak intensity difference for the observations is $0.16 \,\mathrm{nJ\,m^{-2}\,s^{-1}\,Hz^{-1}\,sr^{-1}}$.

The peak separations of the spatially averaged spectra are $32.18 \kms$ (IRIS), $22.41 \kms$ (3D RT), and $18.58 \kms$ (1.5D RT). The roughly $4 \kms$ higher value of the 3D \ac{rt} vs. 1.5D \ac{rt} computation demonstrates that horizontal velocities can have a visible effect on the peak separation (see also Sect.~\ref{sec:the_role_of_horizontal_velocities} and Appendix~\ref{sec:app-effect-of-horizontal-velocities}). The remaining difference between the model and the observation might be explained by the known correlation between the peak separation and the maximum difference of the vertical velocity along the \ac{los}, as for example shown in Fig.~8 in \citet[][]{2013ApJ...772...90L}, Fig.~5 in \citet{2013ApJ...778..143P}, or Fig.~6 in  \citetalias{2024A&A...692A...6O}. This suggests that the vertical motions of the simulation presented here may be lower than in the real Sun. A more dynamic model atmosphere might be achieved by extending the simulation box in the horizontal directions to allow more interaction between the network elements and the quieter parts. In addition, a higher numerical resolution might lead to an increase in the chromospheric dynamics.

\subsubsection{The role of horizontal velocities in the solution of the \ac{rt} problem }
\label{sec:the_role_of_horizontal_velocities}

In order to identify possible reasons for the discrepancy between the 1.5D and full 3D \ac{rt} spectrum, we now discuss the \mgk spectrum at a single location but with different settings used in the \ac{rt} calculation. We randomly selected a location which showed a large difference in the \ktwo intensities between the full 3D and 1.5D \ac{rt} calculation.
First, we checked whether the different angular quadratures used by the 3D and 1.5D solver have an effect on the spectrum. We therefore computed the spectrum at the selected location assuming a plane-parallel atmosphere but once using the \citet{carlson_1963}-A4 angular quadrature (which is used by the 3D solver) and once with the Gau\ss-Legendre quadrature based on five angles (which is used by the 1.5D solver). As can be seen in Fig.~\ref{fig:fig-quadrature-and-vh} (a), there is no significant difference between the two computations, demonstrating that both specific quadratures are compatible in this example.

Next, we tested whether the horizontal velocity field can have an impact on the resulting spectrum when the \ac{rte} is solved on the plane-parallel atmosphere. In the RH1.5D \citep{2001ApJ...557..389U,2015A&A...574A...3P} code, that we used in \citetalias{2024A&A...692A...6O}, and in the 1.5D solver of Multi3D the horizontal velocities are not taken into account. We therefore used the 3D solver of Multi3D but applied it to a plane-parallel atmosphere constructed from the selected location. By doing so, we can directly study the effect of the horizontal velocity field on the emergent spectrum in a de facto 1D \ac{rt} computation. As shown in Fig.~\ref{fig:fig-quadrature-and-vh} (b) the line profile is significantly affected by the horizontal velocities. The line is broadened and the \ktwo peak intensity is reduced. In panel (c) we compare the spectrum from the plane-parallel atmosphere (including horizontal velocities) with the full 3D \ac{rt} solution that takes all inhomogenities of the 3D atmosphere into account. The line shape of the full 3D solution is even broader, and the peak intensities are slightly more reduced. Interestingly, the solution of the plane-parallel atmosphere with horizontal velocities is in this example closer to the full 3D solution than to the 1D \ac{rt} computation without horizontal velocities (panel b). Especially the \ktwor peak results in nearly the same intensity. In Appendix~\ref{sec:app-effect-of-horizontal-velocities}, we present the net effect of the horizontal velocities on the spatially averaged spectrum and the statistical distributions of spectral line parameters by comparing the full 3D computation with a 3D computation where the horizontal velocities are set to zero. The results shown in Fig.~\ref{fig:appendix-role-of-vh-av-spectrum} and Fig.~\ref{fig:rr-quantitatice-comparison-and-effect-of-vh} are similar: the horizontal velocities reduce the peak intensities and slightly increase the peak separation.

From this experiment, we conclude that the horizontal velocity field can have a significant impact on the resulting line shape, even when the spectra are computed in a plane-parallel atmosphere. This suggests the observed difference between the ''standard`` 1.5D \ac{rt} calculation, that is without horizontal velocity field, and the full 3D \ac{rt} solution as visible in Fig.~\ref{fig:average_spectrum} might therefore partly be a result of the dynamic horizontal velocity field in the \ac{muramche} \ac{en} model. This finding is similar to the result of \citet{2021ApJ...909..183J} who studied the effect of the horizontal velocity field in the forward modeling of the \ion{Ca}{i}$\,4227\AA$ line. We note that we calculated all presented spectra in Fig.~\ref{fig:fig-quadrature-and-vh} with full \ac{prd}. In the \ac{crd} approximation, a similar effect is visible. The effect in \ac{crd} is however reduced since \ac{crd} already broadens the line and lowers the \ktwo intensities \citep[see \eg,][Fig.~10]{2017A&A...597A..46S}.

\subsubsection{Intensity images of the observation and the synthesized snapshot}
\label{sec:intensity_images_observation_and_synthesis}
\begin{figure*}

\includegraphics[width=\textwidth,clip]{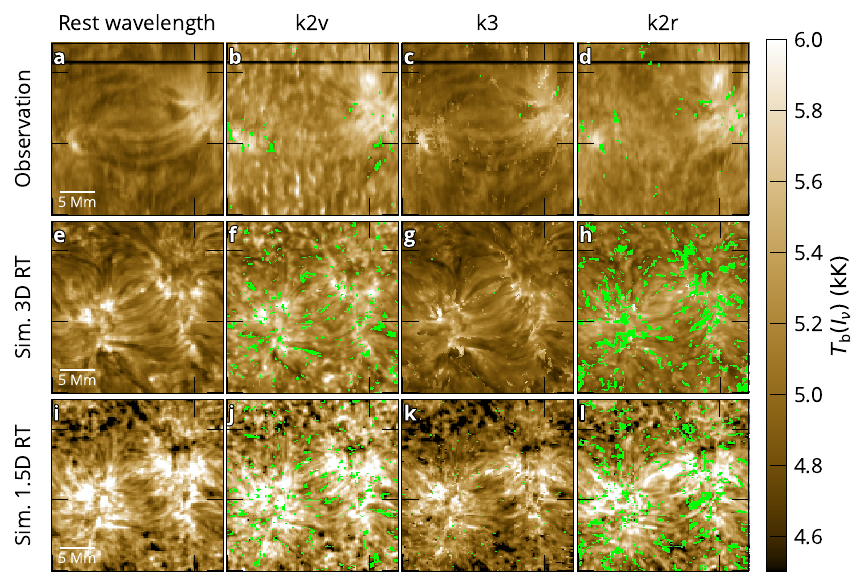}
\caption{Qualitative comparison with observations. We compare three different sets of intensity images. The first row shows the intensity map of the \mgk line from the observation taken at the line-center rest wavelength (panel a), the \ktwov feature (panel b), the \kthree feature (panel c), and the \ktwor feature (panel d). The second row (panels e,f,g, and h) shows the same quantities but for the MURaM-ChE model using 3D RT calculations, and the bottom row (panels i,j,k, and l) when the \ac{rt} problem is solved on a column-by-column approach. The selected \ac{roi} from the observation has the same \ac{fov} as the box size of the simulation. Green pixels indicate spatial locations where no feature could be detected by the peak-finding algorithm. The synthetic spectra were degraded and rebinned to the specifications of the observation dataset.}
 \label{fig:compare_synthesis_with_observation}
\end{figure*}

In this section, we compare the appearance of the intensity images at different spectral features such as the rest wavelength, \ktwo peaks, and \kthree minimum. We aim to demonstrate how 3D \ac{rt} affects the emergent intensity at different locations in the simulation snapshot and how this qualitatively compares with an observation of a bipolar magnetic structure.
Figure \ref{fig:compare_synthesis_with_observation} shows the intensity at the rest wavelength (first column, panels a, e, and i), at the \ktwov feature (second column, panels b, f, and j), the \kthree feature (third column, panels c, g, and k), and the \ktwor feature (fourth column, panels d, h, and l). We compare the observations (top row) with results from our model combined with 3D \ac{rt} (middle row), and computed spectra in the 1.5D \ac{rt} approximation (bottom row). 

The observed bipolar structure shows enhanced intensity above the network polarities in all spectral features. The intensity image at the rest wavelength (panel a) and \kthree feature (panel b) shows fibril-like structures connecting the two polarities. The observed \ktwov image (panel b) shows shock-expansion patterns beneath the bipolar structure, that is, in the upper left and lower right corners of the image. The observed \ktwor image (panel d) shows less sharply defined shock-expansion patterns and is overall fainter. This suggests that the \ktwor peak forms at slightly different heights in the atmosphere than the \ktwov peak. The fact that the \ktwor image (panel d) is more “washed'' out than \ktwov (panel b) might indicate that the red peak forms higher up in the atmosphere.

The intensity images computed with 3D \ac{rt} (middle row, panels e--h) show similar features as the observation, such as connecting fibrilar structures between the polarities. However, they show a higher contrast at the rest wavelength and at the \kthree feature than the observation. The 3D \ac{rt} intensity images taken at the \ktwo peaks (panels f and h) show more structure and higher maximum intensities above the network magnetic fields than the observation (panels b and d). The contrast in the observed images might be lower due to the presence of stray light.

The intensity images from the 1.5D \ac{rt} computations show two dominant differences compared with the 3D \ac{rt} computation. First, the intensity above the network fields is much higher, and in the quiet region (upper left and lower right edge of the images), there exist much fainter regions. The second difference is that at the rest wavelength (panel i) and at the \kthree feature (panel k), the fibrilar structures are hardly visible, whereas in the 3D \ac{rt} computation (\eg, panel g), they are clearly visible.

\subsubsection{Statistical comparison with the observation}

\begin{figure}
\centering
\includegraphics[width=\hsize,clip]{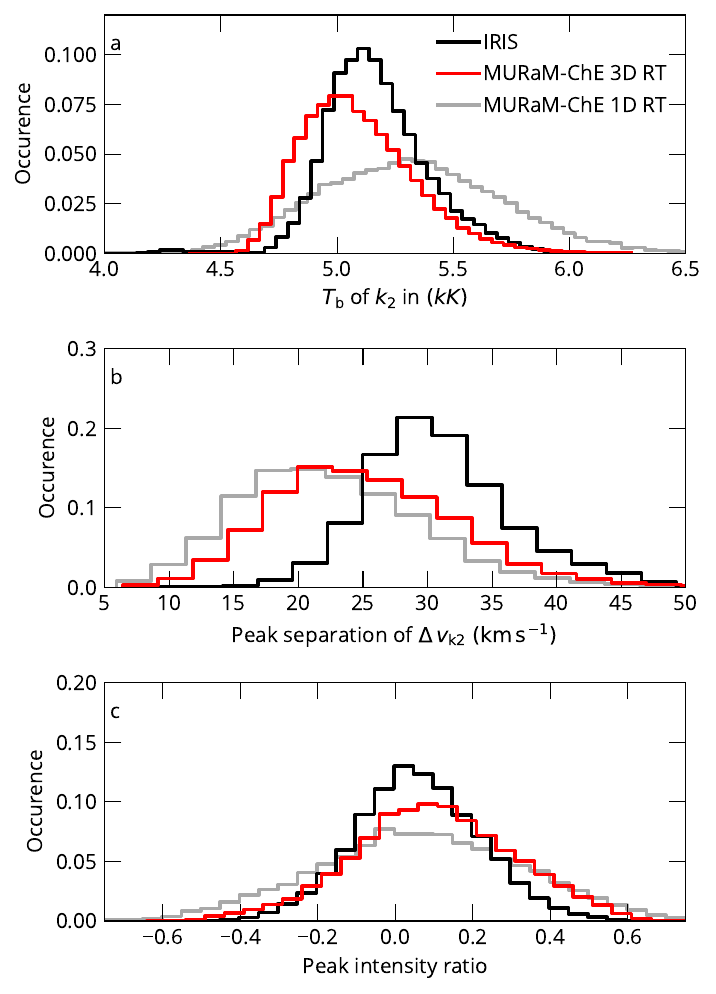}
\caption{Statistical comparison with observations. In panel (a) we show distributions of the \ktwo peak brightness temperature, in panel (b) the distribution of \ktwo peak separation, and in panel (c) the distribution of the \ktwo peak intensity ratio. We show data from the \iris observation in black and data from the synthetic spectra computed by 3D \ac{rt} in red, and 1.5D \ac{rt} in grey after degradation to instrumental conditions.}
 \label{fig:fig3-compare_quantitiatively_synthesis_with_observation}
\end{figure}

Comparing an average spectrum can be overly influenced by the high-intensity pixels. We therefore study here the distribution of the spectral line features. In Fig. \ref{fig:fig3-compare_quantitiatively_synthesis_with_observation}, we show distributions of peak brightness temperature (panel a), \ktwo peak separation (panel b), and \ktwo peak intensity ratio (panel c) of the computed spectra and the observations. The peak brightness temperature of the observation is on average $5.19 \, \mathrm{kK}$ with minimum values of $\approx 4.75 \, \mathrm{kK}$ and maximum values of $\approx 5.9 \, \mathrm{kK}$. The distribution has a Gaussian shape with an extended tail towards higher temperatures, similar to a log-normal distribution. Such log-normal distributions have also been found for the brightness at other wavelengths \citep{2000A&A...362..737P}. The distribution of the peak brightness temperatures from the 3D \ac{rt} computations has a similar shape as the observation but slightly shifted towards lower temperatures. The mean is $5.12 \, \mathrm{kK}$, that is approximately $100 \, \mathrm{K}$ lower than in the observed distribution. The minimum and maximum values are here $\approx 4.6 \, \mathrm{kK}$ and $6.25 \, \mathrm{kK}$. The highest peak brightness temperature has, however a very low occurrence. The results of the spectra computed with 1.5D \ac{rt} are different in mainly two ways. First, the distribution is broader extending to lower ($\approx 4.4 \, \mathrm{kK}$) and higher ($\approx 6.5 \, \mathrm{kK}$) brightness temperatures. Second, the average brightness temperature of $5.34 \, \mathrm{kK}$ from the 1.5D \ac{rt} computation is higher than the 3D \ac{rt} computation and the observation. This result demonstrates that in the \ac{muramche} simulation peak brightness temperatures higher than $\approx 5.25 \, \mathrm{kK}$ are overestimated in the 1.5D \ac{rt} approach.

The observed peak separation distribution has an average value of $32.24 \kms$ and shows an approximately Gaussian shape, however, with slightly more occurrence of higher values. The smallest observed values are $\approx 17 \kms$ and the largest are $\approx 45 \kms$. The peak separation distribution of the 3D \ac{rt} spectra shows a slightly higher average of $26.62 \kms$ compared to the 1.5D \ac{rt} spectra, which are on average $24.01 \kms$. The value of the 1.5D \ac{rt} spectra is $\approx 0.6 \kms$ larger than what we found in \citetalias[][(Sect.~3.2.3.)]{2024A&A...692A...6O}. This could be because of a difference in the codes, the smaller atom model, or the fact that we considered only every other column in the atmosphere for the \ac{rt} computations here. The difference is however small compared for example to the wavelength resolution of \ac{iris}, which is $2.4 \kms$. The peak-separation distribution of the 3D \ac{rt} spectra seems to be shifted by $2-3 \kms$ toward higher values. This might be because the spectra computed in 3D \ac{rt} have a smaller number of small peaks such that the peak separation is better defined. It may also be partly a result of horizontal velocities which are taken into account in the full 3D \ac{rt} synthesis, but not in the 1.5D \ac{rt} synthesis (see Sect.~\ref{sec:the_role_of_horizontal_velocities}).

We computed the peak intensity ratio (Eq.~\ref{eq:paper-ii-peak-intensity-ratio}) for the observation and the computed spectra. The corresponding distributions are shown in panel (c). 
The average values are $0.09$ (\ac{iris}), $0.09$ (1.5D \ac{rt}), and $0.12$ (3D \ac{rt}). These values compare to the ones estimated from the spatially averaged spectrum (Sect.~\ref{sec:results_average_spectra}). The peak intensity ratio is again largest for the 3D \ac{rt} spectra. As described in Sect.~\ref{sec:results_average_spectra}, the difference between the 1.5D \ac{rt} and 3D \ac{rt} values might originate from the fact that in the peak intensity ratio, the differences are normalized by the sum of the intensities. Irrespective of whether the spectra are computed with 1.5D or 3D \ac{rt}, the model produces broader distributions of the peak intensity ratio than in the observation.

\subsection{Correlations between spectral line properties and the atmosphere}
In this second part of our results, we study correlations between spectral line properties and atmospheric properties. The diagnostic potential of such correlations has been presented by \citet{2013ApJ...772...90L} and \citet{2013ApJ...778..143P}. Our aim is here to test whether similar correlations can be found in a different model of the solar chromosphere and how strongly they depend on the model. We begin with a discussion of intensity images at the \ktwov, \ktwor, and \kthree features of the \mgk line. We compare the intensity images to maps of the corresponding formation heights as well as vertical velocity, temperature, and density at the formation height. We then present correlations between the Doppler shifts of the \ktwo and \kthree features and the vertical velocities at their formation heights. We further study the relationship between the peak intensity ratio and the average velocity between the formation height of the \ktwo peaks and the \kthree feature. Finally, we test the potential of the \ktwo peak to infer chromospheric temperatures. 

\subsubsection{Images of synthetic intensity and atmospheric properties}
\label{sec:intensity_images_and_atmosphere}

\begin{figure*}
\centering
\includegraphics[width=0.79\hsize,clip]{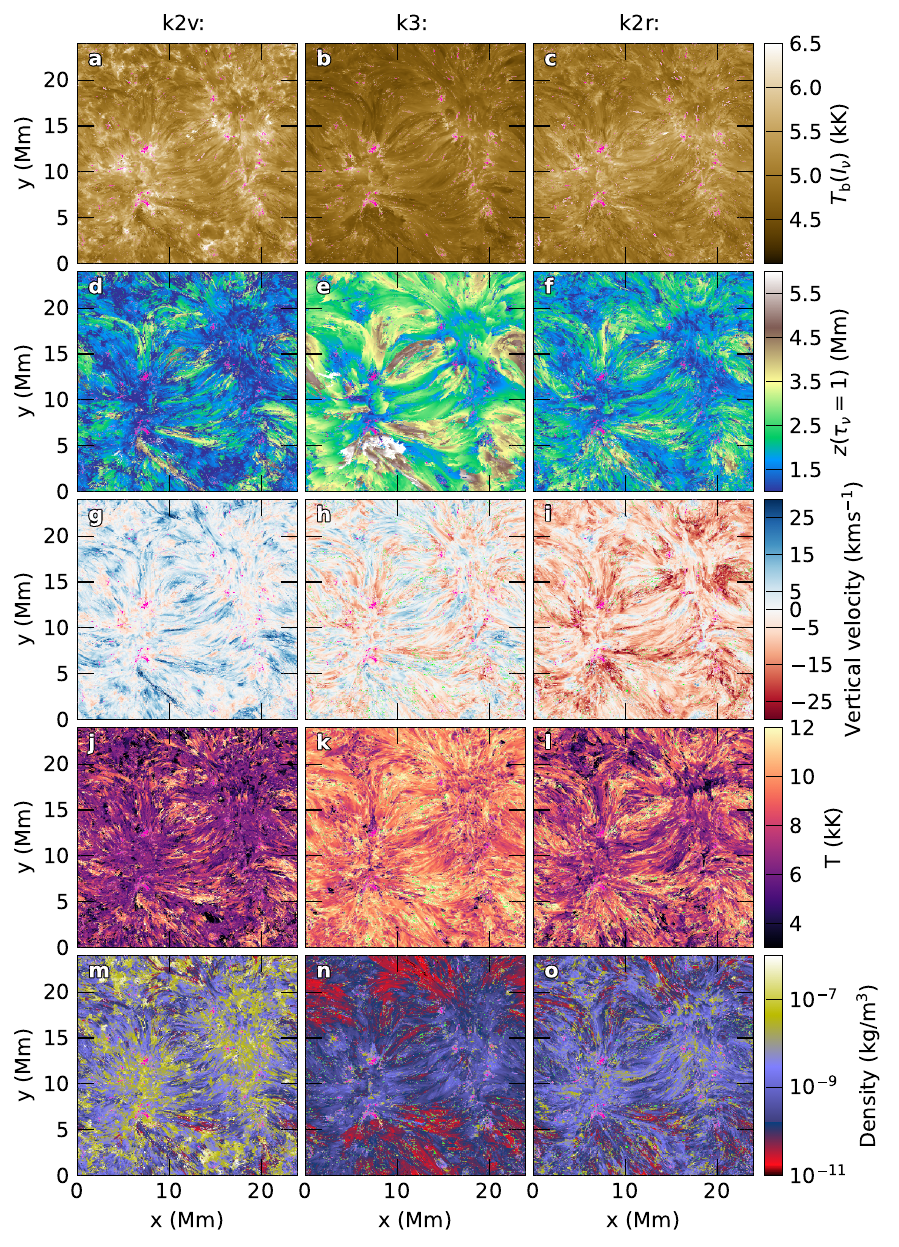}
  \caption{Atmospheric properties at the formation height of the \mgk spectral line features. Intensity maps (first row, panels a--c), formation heights (second row, panels d--f), vertical velocity maps (third row, panels g--i), temperature (fourth row, panels j--l), and density (fifth row, panels m--o). We show these quantities for the \ktwov (left column, panels a,d,g,j, and m), \kthree (middle column, panels b,e,h,k, and n), and \ktwor (right column, panels c,f,i,l, and o) spectral features. The synthetic spectra correspond to the snapshot \snapfourNineNine. Pink-colored pixels indicate where the respective spectral feature could not be detected by the peak-finding algorithm.}
     \label{fig:fig2_intensity_maps}
\end{figure*}

The formation of spectral features of the \mghk lines is known to depend on both the temperature and velocity field over a wide range of heights in the chromosphere. To demonstrate these relationships, we present in Fig.~\ref{fig:fig2_intensity_maps} intensity maps, formation heights, the vertical velocity, temperature, and density at the height of formation $\tau_{\nu} = 1$ of the spectral feature in question (given at the top of the figure). 

We find the \ktwov peak shows different formation properties compared to the \ktwor peak. The intensity image of \ktwov (panel a) shows more shock expansion patterns throughout the whole simulation domain than in the \ktwor intensity image (panel c). Shock expansion patterns are web-like structures, similar to reversed granulation, of enhanced intensity of $T_{\mathrm{b}} > 6 \, \mathrm{kK}$. They are more clearly visible in the quiet regions of the simulation, that is at $y\leq 5 \Mm$ and $y\geq 17 \Mm$. An example is visible at $(x,y) = (5 \Mm, 22\Mm)$ in panel (a). While similar structures exist in the \ktwor image (panel c), they are fainter and show different shapes. The \ktwor intensity image rather shows fibrillar structures of the upper chromosphere as traced by the \kthree feature (panel b). This suggests \ktwov forms lower in the atmosphere than \ktwor. This is supported by the formation height maps shown in panel (d, \ktwov) and panel (f, \ktwor). There, it can be seen that at the same location, the \ktwor feature can form up to $\approx 1 \Mm$ higher than \ktwov. In addition, the formation height image of \ktwor shows imprints of the loop-like structure of the upper chromosphere (see \eg, panel e), whereas such structures are hardly visible in panel (d).

The comparison with the atmospheric properties shows additionally that the vertical velocity at the formation height of \ktwov (panel g) shows preferred upflows (\ie $71 \%$ of the pixels) in relatively thin, spatially concentrated structures with an average velocity of $\langle v_z({h(\tau_{\mathrm{k2v}}=1))}\rangle=3.7 \kms$. In contrast, the vertical velocity at the formation height of \ktwor (panel i) traces downflows (in $86\%$ of the pixels) in spatially more expanded structures with an average velocity of $\langle v_z({h(\tau_{\mathrm{k2r}}=1))}\rangle=-7.3 \kms$. The vertical velocity map traced by the formation height of the \kthree feature shows a rather balanced appearance of up- ($43\%$) and downflows ($57\%$). The maps of temperature and density similarly highlight differences in the formation of \ktwov and \ktwor. The temperature map of \ktwov (panel j) shows lower atmospheric temperatures (on average $6.8 \, \mathrm{kK}$) but in regions of higher density, on average $3.58 \times 10^{-8} \, \mathrm{kg\, m^{-3}}$. Clearly visible in panel (m) is the web-like shock pattern in yellow color at densities of $\approx 10^{-7} \, \mathrm{kg\, m^{-3}}$. In contrast, the temperature at the formation height of \ktwor is higher (on average $8 \, \mathrm{kK}$) and the density is lower (on average $6.7 \times 10^{-9} \, \mathrm{kg\, m^{-3}}$). The higher densities at the formation of \ktwov lead to a stronger coupling between the radiation field and the local gas temperature and thus result in higher peak intensities.

\subsubsection{Correlation between Doppler shifts and vertical velocity at the formation height}
\label{sec:correlation_doppler_shifts_and_atmospheric_velocities}

\begin{figure*}
\centering

\includegraphics[width=\textwidth,clip]{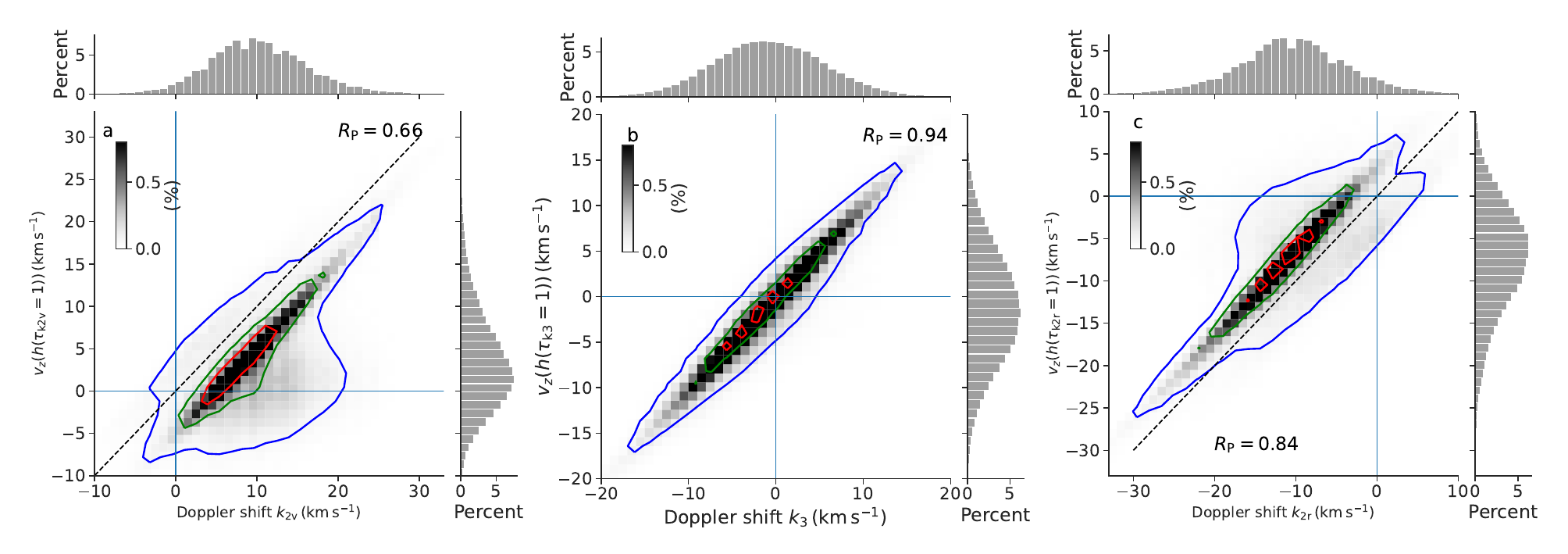}

   \caption{Correlations between Doppler shift of spectral features of \mgk and the vertical velocity in the atmosphere. Panel (a): Correlation between Doppler shift of \ktwov and the vertical component of the velocity at the formation height. Panel (b) shows a similar relation, but for \kthree. Panel (c) is the same as panel (a) but for \ktwor. The Pearson correlation coefficient $R_{\mathrm{P}}$ is given in each panel. The blue solid lines indicate ``$x=0$'' and ``$y=0$''. The black dashed line indicates ``$x=y$''. The red, green, and blue contours enclose $25\%, 50\%,$ and $90\%$ of the data. }
     \label{fig:velocity-diagnostics}
\end{figure*}

We now study the diagnostic potential of the \ktwo and \kthree Doppler shift in more detail. In Fig. \ref{fig:velocity-diagnostics} we show correlations between the Doppler shift and the vertical component of the atmospheric velocity for \ktwov (panel a), \kthree (panel b), and \ktwor (panel c). The correlations based on the \ktwo features indicate a correlation such that the Doppler shift of \ktwov (\ktwor) correlates with upflows (downflows) at the corresponding height of formation in the atmosphere. The Pearson correlation coefficients are $R_{\mathrm{p}}=0.66$ (\ktwov) and $R_{\mathrm{p}}=0.84$ (\ktwor). The correlations of \ktwov and \ktwor are offset toward positive  (\ktwov) and negative (\ktwor) Doppler velocities due to the location of the spectral features with respect to the line core. There are regions in panels (a) and (c) where the spectra show a large Doppler shift but the vertical velocity component in the atmosphere is small. These profiles are broad because of a peak in the temperature stratification in the lower atmosphere \citepalias[see][Sect.~3.2.4.]{2024A&A...692A...6O}. 

The correlation between the Doppler shift of \kthree, and the atmospheric velocity at the formation height is tight with a Pearson correlation coefficient of $R_{\mathrm{P}}=0.94$. This coefficient is slightly lower than the coefficient found by \citet{2013ApJ...772...90L}, which was $R_{\mathrm{P}}=0.99$. For their correlation, the authors computed the line core of \mghk in 3D \ac{rt} but under the \ac{crd} approximation. In \citet[][Fig.~4]{2013ApJ...772...90L} it can also be seen that $\approx 90 \%$ of the data points in the public Bifrost snapshot have vertical velocities within $\pm 10 \kms$ at the formation height of \kthree. In the \ac{muramche} model, the distribution extends to $\approx \pm 16 \kms$ (Fig.~\ref{fig:velocity-diagnostics}b, blue contour), and the scatter is larger. The average vertical velocity at the formation heights of the \kthree feature is $\langle v_z(h(\tau_{\mathrm{k3
}=1}))\rangle = -1.25 \kms$ corresponding to an average Doppler shift of $-1.23 \kms$, which is slightly different from the values obtained from the spatially averaged spectrum. This shows that the Doppler shift of the \kthree feature inferred from the spatially averaged line profile does not necessarily reflect the average of the Doppler shifts of \kthree features from single profiles \citep[see also,][]{2020ApJ...901...32J}.

\subsubsection{Correlation between peak intensity ratio and motions in the atmosphere}

\begin{figure}
\centering
\includegraphics[width=\hsize,clip]{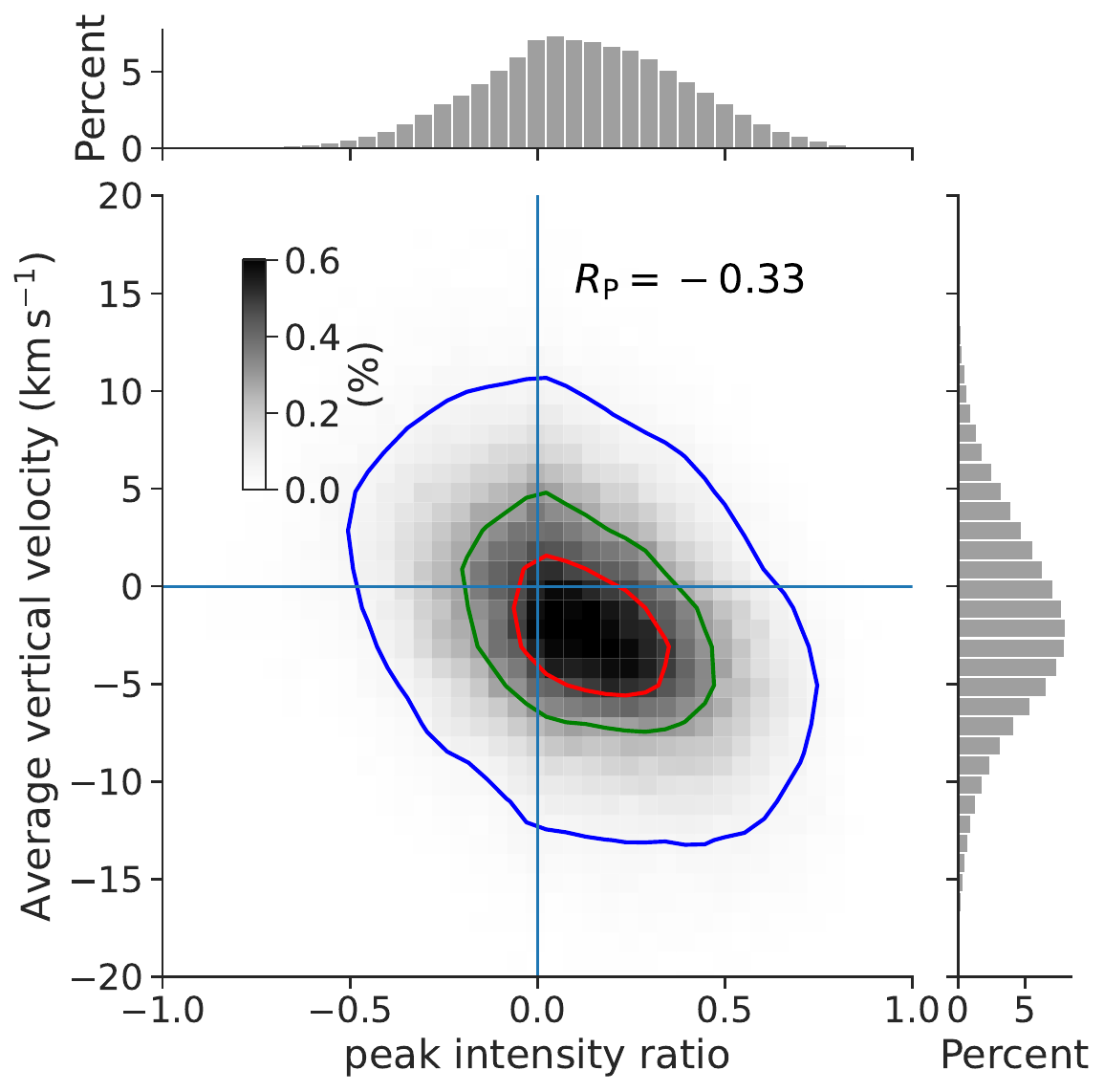}

  \caption{Correlation between peak intensity ratio and average vertical velocity. The average vertical velocity is measured from the minimum formation height of the two \ktwo features and the formation height of the \kthree feature. The red, green, and blue contours enclose $25\%, 50\%,$ and $90\%$ of the data. The blue solid lines indicate ``$x=0$'' and ``$y=0$''.}
     \label{fig:peak-intensity-ratio-average-velocity}
\end{figure}

    The peak intensity ratio (Eq.~\ref{eq:paper-ii-peak-intensity-ratio}) was shown by \citet{2013ApJ...772...90L} to be an indicator of the average vertical velocity in the atmosphere between the formation heights of the \ktwo peaks and \kthree. Due to the large difference between the formation heights of \ktwov and \ktwor we compute the average velocity not from the average \ktwo formation height but from the minimum of the two formation heights of \ktwov and \ktwor \citepalias[see][Eqs.~2 and 3]{2024A&A...692A...6O}. In addition, we exclude cells along the \ac{los} with temperatures of $T_{\mathrm{gas}}>12\,\mathrm{kK}$ as no significant contribution to the intensity is expected from there \citep[see \eg,][Fig.~11]{2012A&A...539A..39C}. As shown in Fig.~\ref{fig:peak-intensity-ratio-average-velocity}, we find a similar correlation to that found by \citet{2013ApJ...772...90L},  that is, $I_{\mathrm{k2v}}>I_{\mathrm{k2r}}$ is correlated with an average downflow in the atmosphere, and vice versa. Our distributions, however, show more scatter and a weaker Pearson correlation coefficient of $R_{\mathrm{P}}=-0.33$, compared to \citet{2013ApJ...772...90L} who found  $R_{\mathrm{P}}=-0.51$. This results from a larger amount of complex spectral line profiles we find in the \ac{muramche} model than compared with the public Bifrost snapshot. Such spectral line profiles show more than two peaks and thus introduce a certain degree of ambiguity for the peak-finding algorithm. The more complex spectral line profiles are a result of the more dynamic atmosphere in the \ac{muramche} model than in the public Bifrost snapshot \citepalias[see \eg,][Fig.~11b]{2024A&A...692A...6O}. Another contribution to the scatter could come from the fact that the topology of the chromosphere in the \ac{muramche} simulation is rather complex, sometimes with hot ($T>12\,\mathrm{kK}$) gas lying between the formation heights of the \ktwo and \kthree features \citepalias[see also][Fig.~7c]{2024A&A...692A...6O}, making a definition of an ‘‘average velocity'' in the considered height range difficult.

\subsubsection{Temperature diagnostics}
Another diagnostic property of the \mghk lines is the correlation between the peak brightness temperature and the atmospheric temperature at the formation heights \citep[see][Fig.~6~e~and~f]{2013ApJ...772...90L}. This correlation, however, tends to be only valid for intensities above a certain minimum brightness temperature of $\approx 5 \, \mathrm{kK}$. These intensities are formed at heights in the chromosphere where the source function is still sufficiently coupled to the local gas temperature.

\begin{figure*}
\centering

\includegraphics[width=16.4cm,clip]{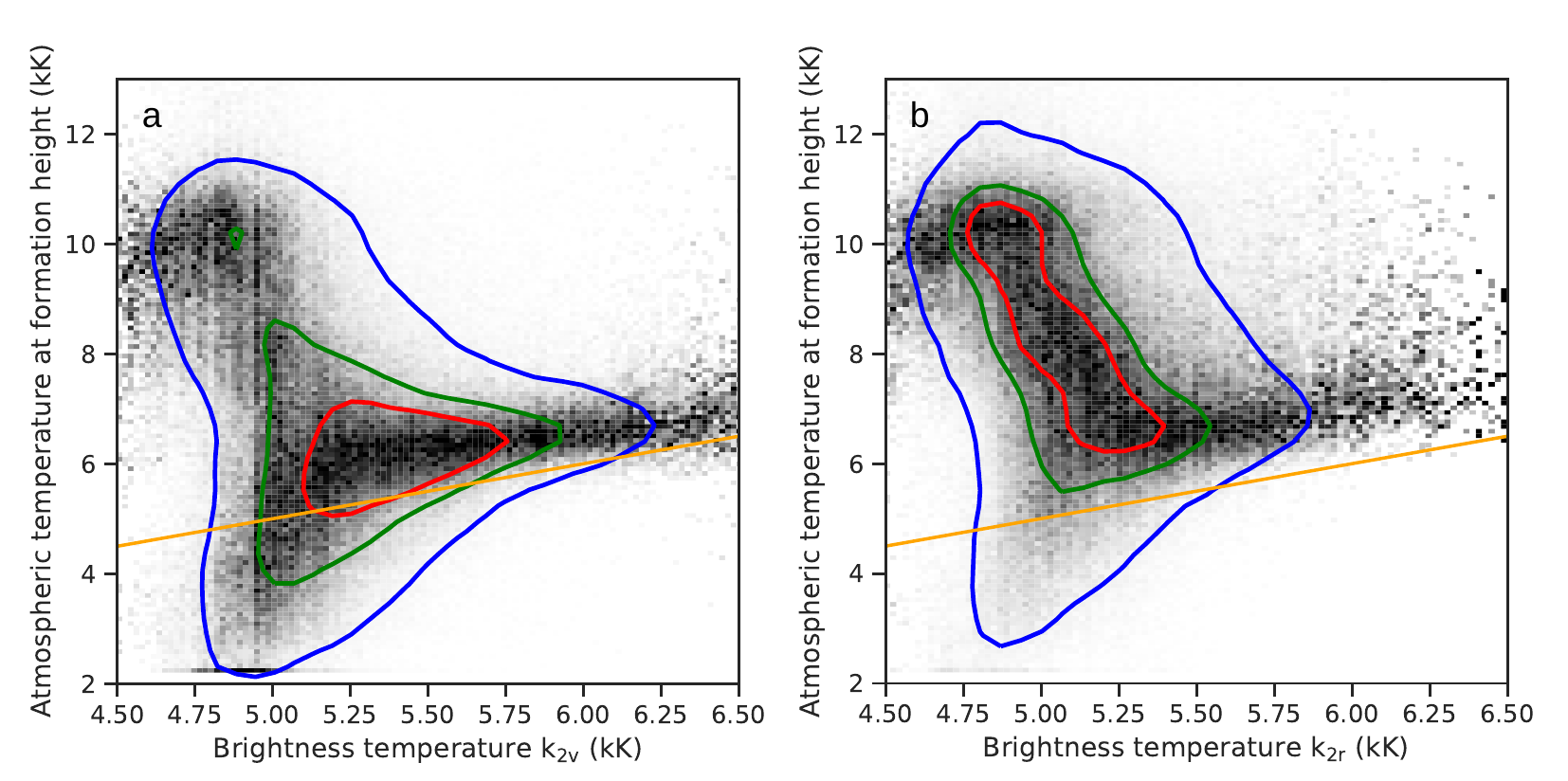}

  \caption{Correlation between peak brightness temperatures and temperature at the formation height in the atmosphere. Panel (a) shows the correlation for the blue peak (\ktwov) and panel (b) shows the correlation for the red peak (\ktwor). The red, green, and blue contours enclose regions of $25\%, 50\%$, and $90\%$ of the data. The orange lines indicate ``$x=y$''. We normalized each column of brightness temperature to the maximum value of atmospheric temperature in that column.}
     \label{fig:correlation_brightness_temperature}
\end{figure*}

We studied the same correlations and present them in Fig.~\ref{fig:correlation_brightness_temperature}. In panel (a) we show the results for the \ktwov feature. It can be seen that there is a correlation between brightness temperatures of $T_{\mathrm{b}} > 5.25 \, \mathrm{kK}$ with the plasma temperature. The Pearson correlation coefficient is, however, only $R_{\mathrm{P}}=0.12$. The reason is that already at the formation height of the \ktwov peak, the line source function can be decoupled from the local gas temperature.  The Pearson correlation coefficient for all datapoints in Fig.~\ref{fig:correlation_brightness_temperature} (a) is  negative ($R_{\mathrm{P}}=-0.15$). 
For $T_{\mathrm{b}} < 5.25 \, \mathrm{kK}$, there is a large variety of atmospheric temperatures, which result in a rather narrow range of brightness temperatures of $T_{\mathrm{b}} \approx 4.75 \, \mathrm{kK}$ to $T_{\mathrm{b}} \approx 5 \, \mathrm{kK}$. These intensities form between the shock-like patterns at a slightly higher formation height. Such scenarios can, for example, be seen in Fig. \ref{fig:fig2_intensity_maps} (a, brightness temperature) and (d, formation height). There is a cluster of datapoints at $T_{\mathrm{gas}}\approx10\,\mathrm{kK}$ where the intensity is decoupled from the local temperature. The maximum atmospheric temperatures of roughly $12\, \mathrm{kK}$ indicate the maximum temperatures where \ion{Mg}{ii} exists dominantly \citep[][Fig.~11]{2012A&A...539A..39C}.

In panel (b), we show the correlation for the \ktwor peak. There is a similar but less tight correlation for $T_{\mathrm{b}} > 5.25 \, \mathrm{kK}$ with a Pearson correlation coefficient of $R_{\mathrm{p}} = 0.06$. The maximum brightness temperatures of \ktwor are lower than those of \ktwov, which was already indicated in the average spectrum (see Fig. \ref{fig:average_spectrum}). For brightness temperatures of $T_{\mathrm{b}} < 5.25 \, \mathrm{kK}$ the correlation is even lower. The overall correlation coefficient is $R_{\mathrm{P}}=-0.21$, which is more negative than for \ktwov. This can be understood from the fact that in our model \ktwor forms higher in the atmosphere than \ktwov, at lower densities and therefore is less suitable to estimate the atmospheric temperature.

Our results compare with the findings of \citet[][Fig.~6e and f]{2013ApJ...772...90L}. These authors found a stronger intensity-temperature correlation for the \ktwov peak than for the \ktwor peak. The actual Pearson correlation coefficients are similarly low. In the public Bifrost snapshot, there is a larger fraction of brightness temperatures below $4.7 \, \mathrm{kK}$. The differences of the temperature stratification between the Bifrost public snapshot and the \ac{muramche} model were discussed in \citetalias[][Fig.~11 (a) and Sect.~3.3]{2024A&A...692A...6O} where the authors found higher temperature in the \ac{muramche} model at the formation heights of \mgk in the chromosphere.

\section{Summary and discussion}
\label{sec:summary_and_conclusion}

\begin{figure}
\centering
\includegraphics[width=\hsize,clip]{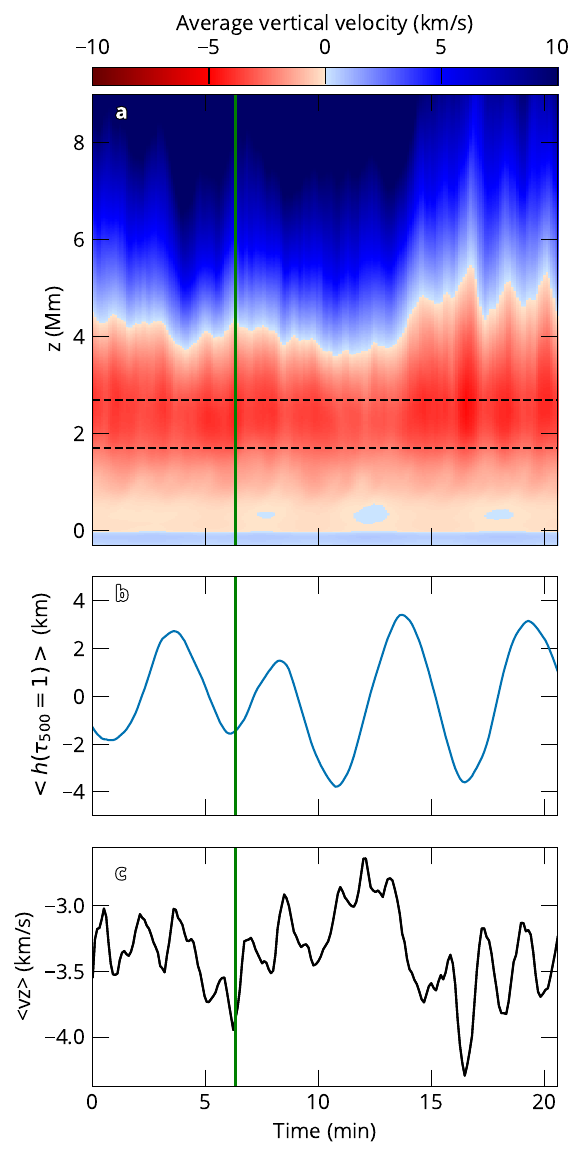}

  \caption{Time-dependent velocity structure in the atmosphere. Panel (a) shows the average velocity as a function of height. Panel (b) shows the time-dependent average height of the \tauUnity surface. Panel (c) shows the average vertical velocity between $z=1.7\Mm$ and $z=2.7\Mm$, which are the average formation heights of the \ktwov feature and \kthree feature in the here presented snapshot. }
     \label{fig:discussion-area-coverage-downflows}
\end{figure}

In this work, we analyzed the spectral line properties of the \mghk lines using 3D \ac{rt} computations in a simulation of a bipolar magnetic feature in the solar atmosphere computed with the \ac{muramche} code. In previous work \citepalias{2024A&A...692A...6O}, we analyzed the \mghk lines of the same simulation snapshot in 1.5D \ac{rt}. Here we discuss the combined results and compare them with previous work.

We determined the spatially averaged spectra, intensity images of the spectral line features, and statistical distributions of line parameters from 1.5D \ac{rt} and 3D \ac{rt} computations and from a dataset recorded by the IRIS spacecraft and compared the results from these three sources. We also compared with previous results for the same spectral lines obtained from Bifrost simulations. The \ac{fov} of the observation was chosen to include a similar bipolar magnetic feature as in the simulation. 

We found that the distribution of the \ktwo and \kthree intensities cover a much wider range and in general display larger values for the 1.5D \ac{rt} computation than for the 3D \ac{rt} computation. Therefore, for the spatially averaged spectrum these intensities are up to $72\%$ higher in the profile obtained via 1.5D \ac{rt} than in the profile from 3D \ac{rt} computations. Thus, in the \ac{muramche} model, the effects of 3D \ac{rt} are important not only in the \kthree minimum but in the whole line inside the \kone minima. While \citet[][Fig.~10]{2017A&A...597A..46S} found similar results for computations with the public Bifrost snapshot, the difference between the 1.5D \ac{rt} and 3D \ac{rt} spatially averaged spectra was much smaller (approximately $12\%$ for \ktwor). In the 1.5D \ac{rt} \ac{muramche} computation, the \ktwo intensities are most strongly overestimated above the network fields. There, the formation heights of \ktwo are relatively low ($<1.5\Mm$, see for example  Fig.~\ref{fig:fig2_intensity_maps}d and f), such that higher densities lead to a stronger coupling between the radiation field to the local temperature and, thus higher emergent intensities. In the plane-parallel \ac{rt} approximation, the horizontal gradients in density and temperature that connect the network-dominated region with the quieter part of the simulation are neglected, which may partly lead to the overestimation. 

In addition, in Fig.~\ref{fig:fig-quadrature-and-vh} and Appendix \ref{sec:app-effect-of-horizontal-velocities}, we demonstrate that the horizontal velocities can have a significant effect on the line shape and especially the \ktwo intensities. This is in agreement with \citet{2021ApJ...909..183J} who found that horizontal velocities in 1.5D improve the \ac{rt} calculation towards the full 3D solution. While these authors studied the \ion{Ca}{i} $4227\AA$ in the \ac{crd} approximation, we found that for the \mghk lines the effect is even more strongly visible in the \ac{prd} computation. The relative difference between the 3D vs. 1.5D spatially averaged spectra, which is larger in the \ac{muramche} \ac{en} than for the public Bifrost snapshot \citep[see][Fig.~10]{2017A&A...597A..46S}, might therefore also be a result of the different horizontal velocity fields in the two atmosphere models.

The distribution of brightness temperature of the \ktwo peaks in the 3D \ac{rt} computation compares much better to the observed distribution than the 1.5D \ac{rt} case (see Fig.~\ref{fig:fig3-compare_quantitiatively_synthesis_with_observation}a). The strength of the averaged profile is also closer to the observations, although it is now (3D \ac{rt}) slighly underestimated (however, the absolute strength of the profile does depend on the exact amount of magnetic flux in the simulation and the observation, see \citetalias{2024A&A...692A...6O}). 

We found that 3D  \ac{rt} produces a slightly higher peak separation of the spatially averaged spectrum than the 1.5D \ac{rt} computation. This can also be seen in the distributions of the peak separation from single spectra (see Fig.~\ref{fig:fig3-compare_quantitiatively_synthesis_with_observation}b). In Appendix \ref{sec:app-effect-of-horizontal-velocities}, we show that this is also an effect of the horizontal velocities that are taken into account in the full 3D computation. Typical 1.5D \ac{rt} computations instead take only the vertical velocity component into account. A better match between the model and the observation might be achieved by a simulation that includes a similar magnetic bipole, but has a larger horizontal extent to allow more interaction with the quiet part in the simulation. This might lead to a more dynamic atmosphere and possibly higher mass loading, which can have an effect on the line width \citep[see \eg,][]{2023ApJ...944..131H}. In addition, a higher numerical resolution might lead to a more dynamic atmosphere \citep[see \eg,][]{2016A&A...585A...4C,2023ApJ...944..131H}.

The peak intensity ratio is even higher in 3D, making it a worse match to the observations. The too high peak intensity ratio in the 3D \ac{rt} computations could be due to either a too high \ktwov or a too low \ktwor intensity. As we found in Sect.~\ref{sec:intensity_images_and_atmosphere}, the \ktwor peak forms higher in the atmosphere than \ktwov. A too-low \ktwor intensity could be due to multiple reasons. For example, a higher gas density at the \ktwor formation height might increase the intensity through a stronger coupling to the local temperature. As we found in Sect.~\ref{sec:intensity_images_and_atmosphere} and Fig.~\ref{fig:fig2_intensity_maps} (m) and (o), the density at the formation height of \ktwov can be up to two orders of magnitude higher than at the formation height of \ktwor. Another contribution to the peak intensity ratio occurs due to the velocity structure in the atmosphere. Similar to the findings of \citet{1997ApJ...481..500C}, the \ktwov peak is stronger than \ktwor if the local atmosphere above the formation heights of the peaks is moving on average downwards, leading to a positive peak intensity ratio (and vise versa). Such a correlation was found by \citet{2013ApJ...772...90L} for the \mghk lines in the public Bifrost snapshot. The correlation in the \ac{muramche} simulation shows however more scatter. This might be a result of the more dynamic atmosphere, that is, there exists much more variation in vertical velocity along the line of sight. 
While we showed here only the results of one synthesized snapshot, it is not immediately clear whether the spectral line properties, such as peak asymmetry, are time-dependent in the simulation. In \citetalias{2024A&A...692A...6O}, we estimated the spatially averaged spectrum from a lower spatial but higher time resolution computation by synthesizing eight slits in the simulation. There, we found not much variation in the spatially averaged line width, peak intensity, and peak intensity ratio over time.

In addition to this, we checked here whether oscillations in the simulation box affect the flow structure in the chromosphere. In Fig.~\ref{fig:discussion-area-coverage-downflows}, we show the average vertical velocity as a function of height and time (panel a), the average height of the $\tau_{500}=1$ surface as a function of time (panel b), and the average vertical velocity at the approximate formation heights of the \ktwo features (panel c). We find the average \tauUnity height changes periodically within $\approx 6 \, \mathrm{min}$ with an amplitude of $\approx 3$--$4\km$. The oscillation seems to correlate with the average vertical velocity in the lower chromosphere $(z<1\Mm)$. Higher up in the chromosphere at $1.7\Mm \leq z \leq 2.7\Mm$, approximately the heights between where the \ktwov and \kthree features form, the atmosphere shows on average downflows independent of time. In panel (c) we show the average vertical velocity in the atmosphere averaged over the whole $xy$-plane between heights of $z=1.7 \Mm$ and $z=2.7\Mm$. It can be seen that these velocities are always negative between $\approx -2.6\kms$ and  $\approx-4.3\kms$. Therefore, in accordance with Fig.~\ref{fig:peak-intensity-ratio-average-velocity} and the results from \citet[][Fig.~8g]{2013ApJ...772...90L}, it might be that there is a positive peak asymmetry ratio in the spatially averaged spectrum at almost any snapshot in the simulation. We note however, as discussed above, that other atmospheric parameters play an important role in line formation, in addition to the velocity field.

We found that correlations between spectral line properties and the atmosphere exist similarly to the findings of \citet{2013ApJ...772...90L} and \citet{2013ApJ...778..143P}. We could confirm that a tight correlation between the vertical velocity at the formation height of \kthree and the Doppler shift of this spectral feature exists. In addition, we found that the Doppler shifts \ktwov (\ktwor) correlate preferrentially with upflows (downflows). This means the \ktwov and \ktwor peaks of the same ray may probe significantly different layers of the atmosphere. In particular, we found \ktwov traces lower-lying regions of higher density, probably due to shock compression. Whereas \ktwor forms higher up at lower densities and on average at larger velocities.

The \ktwo peak intensity as a temperature diagnostic is only partly valid in the \ac{muramche} simulation. We found, in general agreement with  \citet{2013ApJ...772...90L}, that for \ktwov peak brightness temperatures $>5.25\,\mathrm{kK}$, there seems to be a good correlation. The Pearson coefficient is, however, relatively low ($0.12$).  In the case of \ktwor, this correlation is weaker with a Pearson correlation coefficient of $0.06$, which can be understood by our findings that \ktwor forms higher up in the chromosphere, where \ac{nlte} effects become more important.

As a general result, we find the \ac{muramche} model is able to produce a close match to the observed line width in 1.5D \ac{rt} and 3D \ac{rt} at a relatively moderate resolution of $23.4 \km$ (horizontal) and $20 \km$ (vertical). Whether a higher resolution will improve the match with the observation needs to be shown in the future.

\section{Conclusion}
\label{sec:conclusion}
We performed 3D \ac{rt} computations with the Multi3D code in an \ac{en} simulation that was computed with the \ac{muramche} code. We confirmed that the effects of horizontal \ac{rt} must be taken into account to accurately model the \mghk lines. The difference between 1.5D \ac{rt} and 3D \ac{rt} was even more pronounced than in previous studies with the public Bifrost snapshot. In addition, we confirmed the diagnostic potential of the \mgk line to constrain the velocity structure of observations. We found that \ktwov and \ktwor trace different features in the chromosphere, which might help to better interpret observed spectra. While differences still exist between the 3D \ac{rt} computations of the \mghk lines and the observation, our results demonstrate progress in numerical modeling of the chromosphere. 

\begin{acknowledgements}

 We thank the anonymous referee for comments and suggestions that improved the quality of this paper. This work was supported by the International Max-Planck Research School (IMPRS) for Solar System Science at the University of Göttingen. This work received funding from the European Research Council (ERC) under the European Union's Horizon 2020 research and innovation programme (grant agreement No. 101097844-project WINSUN). This research has received financial support from the European Union’s Horizon 2020 research and innovation program under grant agreement No. 824135 (SOLARNET). This work was supported by the Deutsches Zentrum f{\"u}r Luft und Raumfahrt (DLR; German Aerospace Center) by grant DLR-FKZ 50OU2201. We highly appreciate the computing resources provided by the HPC systems Raven, Cobra, and Viper at the Max Planck Computing and Data Facility. 
\end{acknowledgements}
\bibliographystyle{aa}
\bibliography{bib-ads}
\begin{appendix}
\section{The effect of horizontal velocities on the \mghk lines}
\label{sec:app-effect-of-horizontal-velocities}

\begin{figure}
\includegraphics[width=\hsize, clip]{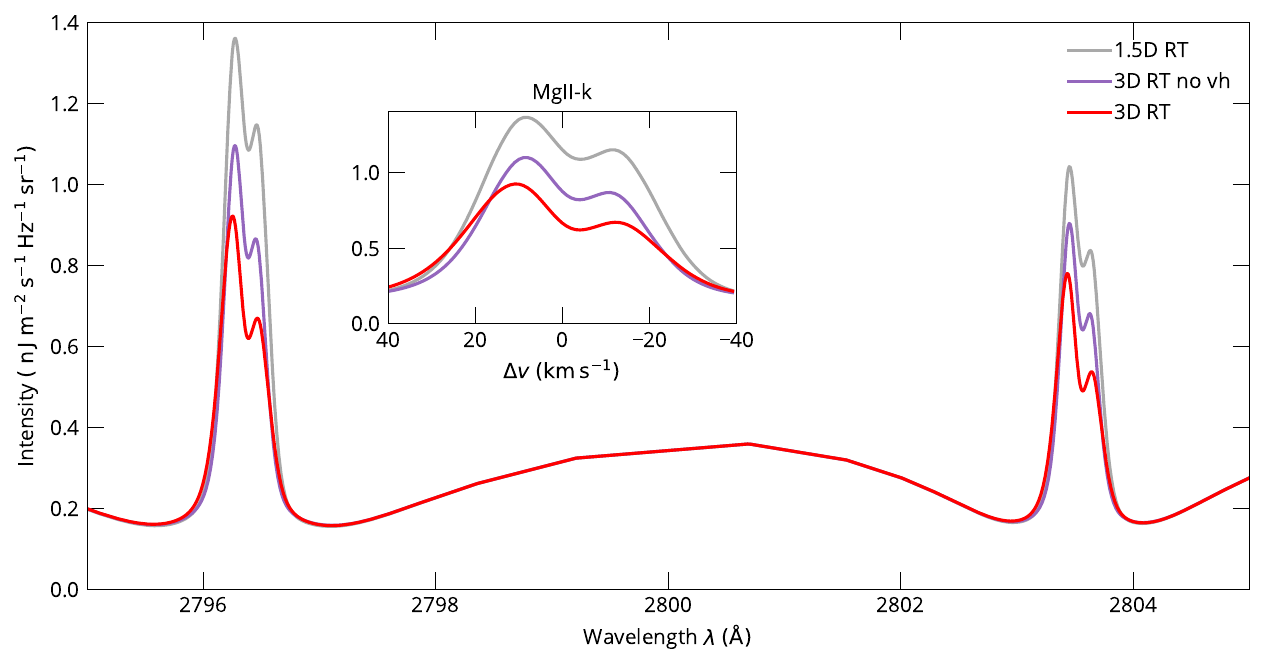}
  \caption{The effect of horizontal velocities on the spatially averaged line profile. We show the results of three different computations. In grey: 1.5D \ac{rt}, computed without horizontal velocities. In purple: 3D \ac{rt}, but without horizontal velocities. In red: 3D \ac{rt} computation that takes the full atmosphere into account.}
     \label{fig:appendix-role-of-vh-av-spectrum}
\end{figure}

\begin{figure}
\includegraphics[width=\hsize,clip]{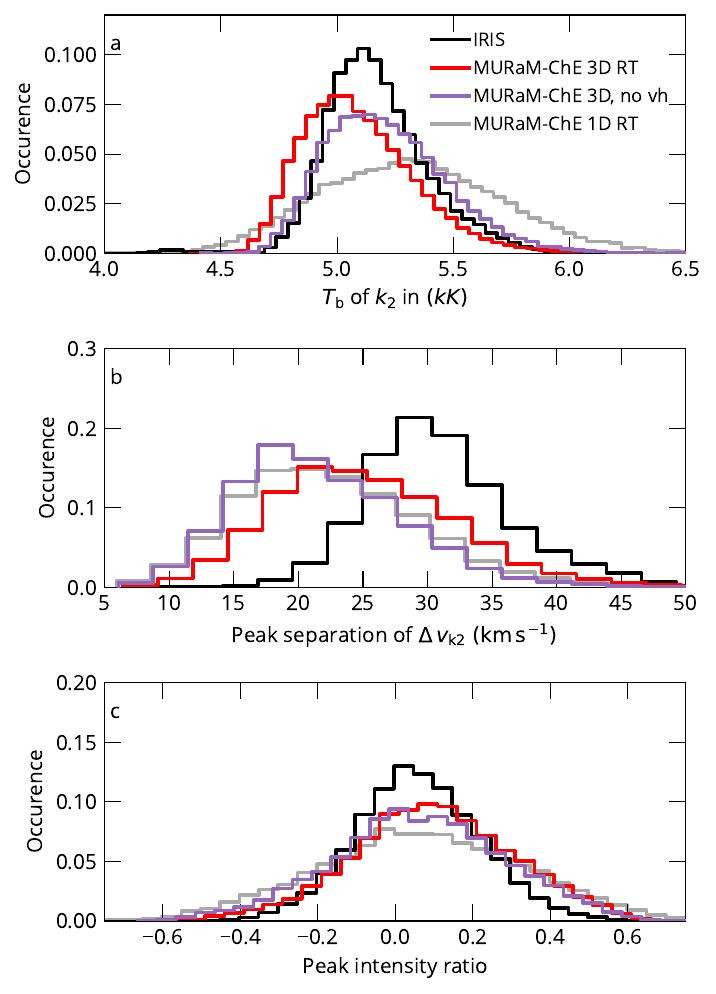}
 \caption{Effect of horizontal velocities on the distributions of spectral line parameters. The layout is identical to Fig.~\ref{fig:fig3-compare_quantitiatively_synthesis_with_observation} in the main text. Panel (a) shows the distributions of peak brightness temperature, panel (b) the distributions of peak separation, and panel (c) the distribution of the peak intensity ratio. In addition, we show in purple the results from the 3D \ac{rt} computation where the horizontal velocities are set to zero. }
     \label{fig:rr-quantitatice-comparison-and-effect-of-vh}
\end{figure}

In this appendix, we demonstrate the effect of the horizontal velocities on the spatially averaged line profile and on single profiles computed at the disk center. In Fig.~\ref{fig:appendix-role-of-vh-av-spectrum}, we show spatially averaged profiles resulting from the 1.5D \ac{rt}, 3D \ac{rt} without horizontal velocities, and the full 3D \ac{rt} computation. It can be seen that the difference between the 1.5D \ac{rt} and the full 3D \ac{rt} computation is partly due to inhomogeneities in atmospheric quantities such as temperature or density, but also that the horizontal velocity field plays an important role. The 3D \ac{rt} computation without horizontal velocities results in a slightly smaller peak separation and higher peak intensities than the full 3D computation. The same results can be seen by looking at the statistical distributions of the spectral line parameters as presented in Fig.~\ref{fig:rr-quantitatice-comparison-and-effect-of-vh}, which is similar to Fig.~\ref{fig:fig3-compare_quantitiatively_synthesis_with_observation} in the main text, but additionally contains the results of the 3D \ac{rt} computation with horizontal velocities set to zero. The distribution of the peak brightness temperature is shifted towards higher values in the 3D \ac{rt} computation without horizontal velocities compared to the 3D \ac{rt} computation taking the full atmosphere into account. The distribution of the peak separation computed without horizontal velocities is similar to the 1.5D \ac{rt} computaton, that does by default not take horizontal velocities into account. Finally, the distribution of the peak intensity ratio is only slightly affected by the horizontal velocities.

\end{appendix}
\end{document}